\documentclass[11pt]{revtex4-1}

\usepackage{graphicx}

\begin{document}

\title{QUANTUM TUNNELLING FOR HAWKING RADIATION FROM BOTH STATIC AND DYNAMIC BLACK HOLE}

\author{Subenoy Chakraborty\footnote {schakraborty.math@gmail.com}}
\author{Subhajit Saha\footnote {subhajit1729@gmail.com}}
\affiliation{Department of Mathematics, Jadavpur University, Kolkata-700032, West Bengal, India.}

\begin{abstract}
The paper deals with Hawking radiation from both a general static black hole and a non-static spherically symmetric black hole. In case of static black hole, tunnelling of non-zero mass particles are considered and due to complicated calculations, quantum corrections are calculated only upto first order. The results are compared with those for massless particles near the horizon. On the other hand, for dynamical black hole, quantum corrections are incorporated using Hamilton-Jacobi method beyond semi-classical approximation. It is found that different order correction terms satisfy identical differential equation and are solved by a typical technique. Finally, using the law of black hole mechanics, a general modified form of the black hole entropy is obtained considering modified Hawking temperature.\\\\
Keywords: Tunnelling, Quantum Correction, Hawking temperature\\
PACS Numbers: 04.70.Dy, 04.60.Kz, 04.62.+v 

\end{abstract}

\maketitle
\section{INTRODUCTION}

Hawking radiation is one of the most important effects in black hole (BH) physics. Classically, nothing can escape from the BH across its event horizon. But in 1974, there was a dramatic change in view when Hawking [1,2] showed that BHs are not totally black, they radiate analogous to thermal black body radiation. Since then, there has been lots of attraction to this issue and various approaches have been developed to derive Hawking radiation and its corresponding temperature [3-7]. However, in the last decade, two distinct semi-classical methods have been developed which enhanced the study of Hawking radiation to a great extent. The first approach developed by Parikh and Wilczek [8,9] is based on the heuristic pictures of visualisation of the source of radiation as tunnelling and is known as radial null geodesic method. The essence of this method is to calculate the imaginary part of the action for the $s-wave$ emmission (across the horizon) using the radial null geodesic equation and is then related to the Boltzmann factor to obtain Hawking radiation by the relation
\begin{equation}
\Gamma \propto exp\left\lbrace -\frac{2}{h}(Im~S^{out}-Im~S^{in})\right\rbrace =exp\left\lbrace -\frac{E}{T_H}\right\rbrace,
\end{equation}
where $E$ is the energy associated with the tunnelling particle and $T_H$ is the usual Hawking temperature.

The alternative way of looking into this aspect is known as complex paths method developed by Srinivasan {\it et al.} [10,11]. In this approach, the differential equation of the action $S(r,t)$ of a classical scalar particle can be obtained by plugging the scalar field wave function $\phi (r,t)=exp\lbrace{-\frac{i}{\hbar}S(r,t)\rbrace}$ into the Klein-Gordon (KG) equation in a gravitational background. Then, Hamilton-Jacobi (HJ) method is employed to solve the differential equation for $S$. Finally, Hawking temperature is obtained using the "Principle of detailed balance" [10-12] (time-reversal invariant). It should be noted that the first method is limited to massless particles only. Also, this method is applicable to such coordinate system only in which there is no singularity across the horizon.. On the other hand, in complex paths method, the emitted particles are considered without self-gravitation and the action is assumed to satisfy the relatisvistic HJ equation. Here tunnelling of both massless and massive particles are possible and it is applicable to any coordinate system to describe the BH.

Most of the studies [13-18] dealing with the Hawking radiation are connected to semi-classical analysis. Recently, Banerjee {\it et al.} [19] and Corda {\it et al.} [20,21] initiated the calculation of Hawking temperature beyond the semi-classical limit. Mostly, both groups have considered tunnelling of massless particle and evaluated the modified Hawking temperature with quantum corrections.

In the present work, at first we consider a general non-static metric for dynamical BH. HJ method is extended beyond semi-classical approximation to consider all the terms in the expansion of the one particle action. It is found that the higher order terms (quantum corrections) satisfy identical differential equations as the semi-classical action and the complicated terms are eliminated considering BH horizon as one way barrier. We derive the modified Hawking temperature using both the above approaches and are found to be identical at the semi-classical level. Also, modified form of the BH entropy with quantum correction has been evaluated.

Subsequently, in the next section, we consider tunnelling of particles having non-zero mass beyond semi-classical approximation. Due to non-zero mass, the imaginary part of the action cannot be evaluated using first approach, only HJ method will be applicable. Further, the complicated form of the equations involved restricted us to only first order quantum correction.

\section{Method of radial null geodesic: A survey of earlier works}

This section deals with a brief survey of the method of radial null geodesics method [8] considering the picture of Hawking radiation as quantum tunnelling. In a word, the method correlates the imaginary part of the action for the classically forbidden process of {\it s-wave} emmission across the horizon with the Boltzmann factor for the black body radiation at the Hawking temperature. We start with a general class of non-static spherically symmetric BH metric of the form
\begin{equation}
ds^2=-A(r,t)dt^2+\frac{dr^2}{B(r,t)}+r^2d{\Omega _2}^2,
\end{equation}
where the horizon $r_h$ is located at $A(r_h,t)=0=B(r_h,t)$ and the metric has a coordinate singularity at the horizon. To remove the coordinate singularity, we make the following Painleve-type transformation of coordinates:
\begin{equation}
dt\rightarrow dt-\sqrt{\frac{1-B}{AB}}dr
\end{equation}
and as a result the metric (2) transforms to 
\begin{equation}
ds^2=-Adt^2+2\sqrt{A\left(\frac{1}{B}-1\right)}dtdr+dr^2+r^2d{\Omega _2}^2.
\end{equation}
This metric ({\it i.e.,} the choice of coordinates) has some distinct features over the former one, namely\\
$\bullet$ The metric is singularity free across the horizon,\\
$\bullet$ At any fixed some, we have a flat spatial geometry,\\
$\bullet$ Both the metric has the same boundary geometry at any fixed radius.

The radial null geodesic (characterized by $ds^2=0=d{\Omega _2}^2$) has the differential equation (using (3))
\begin{equation}
\frac{dr}{dt}=\sqrt{\frac{A}{B}}\left[\pm 1-\sqrt{1-B(r,t)}\right],
\end{equation}
where outgoing or ingoing geodesic is identified by the + or - sign within the square bracket in equation (4). In the present case, we deal with the absorption of particles through the horizon ({\it i.e.,} + sign only) and according to Parikh and Wilczek [8], the imaginary part of the action is obtained as
\begin{equation}
Im~S=Im \int _{r_{in}}^{r_{out}} p_r dr=Im \int _{r_{in}}^{r_{out}} \int _0^{p_r} dp_r^{'}dr=Im \int _{r_{in}}^{r_{out}} \left\lbrace \int _0^{H}\frac{dH'}{\frac{dr}{dt}}\right\rbrace dr.
\end{equation}
Note that in the last step of the above derivation we have used the Hamilton's equation $\dot{r}=\frac{dH}{dp_r}|_r$, where ($r$,$p_r$) are canoniacl pair. Further, it is to be mentioned that in quantum mechanics, the action of a tunnelled particle in a potential barrier having energy larger than the energy of the particle will be imaginary as $p_r=\sqrt{2m(E-V)}$. For the present non-static BH, the mass of the BH is not constant and hence the $dH^{'}$ integration extends over all values of energy of outgoing particle, from zero to $E(t)$ [22] (say). As we are dealing with tunnelling across the BH horizon, so using taylor series expansion about the horizon $r_h$ we write
\begin{equation}
A(r,t)|_t={\frac{\partial A(r,t)}{\partial r}}|_t(r-r_h)+O(r-r_h)^2|_t,
\end{equation}
\begin{equation}
B(r,t)|_t={\frac{\partial B(r,t)}{\partial r}}|_t(r-r_h)+O(r-r_h)^2|_t.
\end{equation}
So, in the neighbourhood of the horizon, the geodesic equation (4) can be approximated as
\begin{equation}
\frac{dr}{dt}\approx \frac{1}{2}\sqrt{A'(r_h,t)B'(r_h,t)}(r-r_h).
\end{equation}
Substituting this value of $\frac{dr}{dt}$ in the last step of the equation (5) we have
\begin{equation}
Im~S=\frac{2\pi E(t)}{\sqrt{A'(r_h,t)B'(r_h,t)}},
\end{equation}
where the choice of contour for r-integration is on the upper half complex plane to avoid the coordinate singularity at $r_h$. Thus the tunnelling probability is given by
\begin{equation}
\Gamma \sim exp\left\lbrace -\frac{2}{\hbar}Im~S\right\rbrace =exp\left\lbrace -\frac{4\pi E(t)}{\hbar \sqrt{A'B'}}\right\rbrace ,
\end{equation}
which in turn equating with the Boltzmann factor $exp\left\lbrace \frac{E(t)}{T}\right\rbrace$, the expression for the Hawking temperature is 
\begin{equation}
T_H=\frac{\hbar \sqrt{A'(r_h,t)B'(r_h,t)}}{4\pi}.
\end{equation}
From the above expression, it is to be noted that $T_H$ is time dependent.

Recently, a drawback of the above approach has been noted [23-25]. It has been shown that $\Gamma \sim exp\lbrace{-\frac{2}{\hbar}Im~S\rbrace}=exp\lbrace{-\frac{2}{\hbar}Im~\int _{r_{in}}^{r_{out}} p_r dr\rbrace}$ is not canonically invariant and hence is not a proper observable, it should be modified as $exp\left\lbrace -\frac{Im~\oint p_r dr}{\hbar}\right\rbrace$. The closed path goes across the horizon and back. For tunnelling across the ordinary barrier, it is immaterial whether the particle goes from the left to the right or the reverse path. So in that case
\begin{equation}
\oint p_r dr=2\int _{r_{in}}^{r_{out}} p_r dr
\end{equation}
and there is no problem of canonical invariance. But difficulty arises for BH horizon which behaves as a barrier for particles going from inside of the BH to outside but it does not act as a barrier for particles going from outside to the inside. So relation (13) is no longer valid. Also using tunnelling the probability is $\Gamma \sim exp\left\lbrace -\frac{Im~\oint p_r dr}{\hbar}\right\rbrace$, so there will be a problem of factor two in Hawking temperature [24,26].  

Further, the above analysis of tunnelling approach remains incomplete unless effects of self gravitation and back reaction are taken into account. But unfortunately, no general approach to account for the above effects are there in the literature-only few results are available for some known BH solutions [26-29].

Finally, it is worthy to mention that so far the above tunnelling approach is purely semi-classical in nature and quantum corrections are not included. Also this method is applicable for Painleve type coordinates only-one can not use the original metric coordinates to avoid horizon singularity. Lastly, the tunnelling approach is not applicable for massive particles [19].

\section{Hamilton-Jacobi Method: Quantum Corrections}

We shall now follow the alternative approach as mentioned in the introduction, {\it i.e.,} the HJ method to evaluate the imaginary part of the action and hence the Hawking temperature. We shall analyze beyond semi-classical approximation by incorporating possible quantum corrections. As this method is not affected by the coordinate singularity at the horizon so we shall use the general BH metric (2) for convinience.

In the background of the gravitational field described by the metric (2), massless scalar particles obey the Klein-Gordon equation
\begin{equation}
\frac{\hbar ^2}{\sqrt{-g}}\partial [g^{\mu \nu}\sqrt{-g}\partial _{\nu}]\psi =0.
\end{equation}
For spherically symmetric BH, as we are only considering radial trajectories, so we shall consider (t,r)-sector in the spacetime given by the equation (2), {\it i.e.,} we concentrate on two-dimensional BH problems. Using (2), the above Klein-Gordon equation becomes
\begin{equation}
\frac{\partial ^2 \psi}{\partial t^2}-\frac{1}{2AB}\frac{\partial (AB)}{\partial t}\frac{\partial \psi}{\partial t}-\frac{1}{2}\frac{\partial (AB)}{\partial r}\frac{\partial \psi}{\partial r}-AB\frac{\partial ^2 \psi}{\partial r^2}=0.
\end{equation}
Using the standard ansatz for the semi-classical wave function, namely
\begin{equation}
\psi (r,t)=exp\lbrace{-\frac{i}{\hbar}S(r,t)\rbrace},
\end{equation}
the differential equation for the action $S$ is
\begin{equation}
{\left(\frac{\partial S}{\partial t}\right)}^2-AB{\left(\frac{\partial S}{\partial r}\right)}^2+i\hbar \left[\frac{\partial ^2 S}{\partial t^2}-\frac{1}{2AB}\frac{\partial (AB)}{\partial t}\frac{\partial S}{\partial t}-\frac{1}{2}\frac{\partial (AB)}{\partial r}\frac{\partial S}{\partial r}-AB\frac{\partial ^2 S}{\partial r^2}\right].
\end{equation}
To solve this partial differential equation we expand the action $S$ in powers of Planck's constant $\hbar$ as
\begin{equation}
S(r,t)=S_0(r,t)+\Sigma \hbar ^k S_k(r,t),
\end{equation}
with $k$ a positive integer. Note that in the above expansion, terms of the order of Planck's constant and its higher powers are considered as quantum corrections over the semi-classical action $S_0$. Now substituting the ansatz (18) for $S$ into (17) and equating different powers of $\hbar$ on both sides, we obtain the following set of partial differential equations:
\begin{equation}
\hbar ^0:~~{\left(\frac{\partial S}{\partial t}\right)}^2-AB{\left(\frac{\partial S}{\partial r}\right)}^2=0,
\end{equation}
\begin{equation}
\hbar ^1:~~\frac{\partial S_0}{\partial t}\frac{\partial S_1}{\partial t}-AB\frac{\partial S_0}{\partial r}\frac{\partial S_1}{\partial r}+\frac{i}{2}\left[\frac{\partial ^2 S_0}{\partial t^2}-\frac{1}{2AB}\frac{\partial (AB)}{\partial t}\frac{\partial S_0}{\partial t}-\frac{1}{2}\frac{\partial (AB)}{\partial r}\frac{\partial S_0}{\partial r}-AB\frac{\partial ^2 S_0}{\partial r^2}\right]=0,
\end{equation}
\begin{equation}
\hbar ^2:~~{\left(\frac{\partial S_1}{\partial t}\right)}^2+2\frac{\partial S_0}{\partial t}\frac{\partial S_2}{\partial t}-AB{\left(\frac{\partial S_1}{\partial r}\right)}^2-2AB\frac{\partial S_0}{\partial r}\frac{\partial S_2}{\partial r}+i\left[\frac{\partial ^2 S_1}{\partial t^2}-\frac{1}{2AB}\frac{\partial (AB)}{\partial t}\frac{\partial S_1}{\partial t}-\frac{1}{2}\frac{\partial (AB)}{\partial r}\frac{\partial S_1}{\partial r}-\linebreak[0] AB\frac{\partial ^2 S_1}{\partial r^2}\right]=0,
\end{equation}
and so on.

Apparently, different order partial differential equations are very complicated but fortunately there will be lot of simplifications if in the partial differential equation corresponding to $\hbar ^k$, all previous partial differential equations are used and finally we obtain identical partial differential equation, namely
\begin{equation}
\hbar ^k:~~\frac{\partial S_k}{\partial t}=\pm \sqrt{A(r,t)B(r,t)}\frac{\partial S_k}{\partial r},
\end{equation}
for $k=0,1,2...$.

Thus quantum corrections satisfy same differential equation as the semi-classical action $S_0$. Hence the solutions will be very similar. To msolve $S_0$, it is to be noted that due to non-static BHs the metric coefficients are functions of $r$ and $t$ and hence standard HJ method cannot be applied, some generalization is needed. We start with a general metric [22]
\begin{equation}
S_0(r,t)=\int _0^{t}\omega _0(t')dt+D_0(r,t).
\end{equation}
Here $\omega _0 (t)$ behaves as the energy of the emitted particle and the justification of the choice of the integral is that the outgoing particle should have time-dependent continuum energy.

Now substituting the above ansatz for $S_0(r,t)$ into equation (19) and using the radial null geodesic in the usual metric from (2), namely
\begin{equation}
\frac{dr}{dt}=\pm \sqrt{AB}.
\end{equation}
We have,
\begin{center}
$\frac{\partial D_0}{\partial r}+\frac{\partial D_0}{\partial t}\frac{dt}{dr}=\mp\omega _0(t)\frac{dt}{dr}$,
\end{center}
{\it i.e.,}
\begin{center}
$\frac{dD_0}{dr}=\mp \frac{\omega_0 (t)}{\sqrt{AB}}$,
\end{center}
which gives
\begin{equation}
D_0=\mp\omega _0(t)\int _0^{r}\frac{dr}{\sqrt{AB}}.
\end{equation}
Hence the complete semi-classical action takes the form
\begin{equation}
S_0(r,t)=\int _0^{t}\omega _0(t')dt'\mp \omega _0(t)\int _0^{r}\frac{dr}{\sqrt{AB}}.
\end{equation}
Here the - (or +) sign corresponds to absorption (or emmission) particle. As the solution (26) contains an arbitrary time-dependent function $\omega _0(t)$, so a general solution for $S_k$ can be written as
\begin{equation}
S_k(r,t)=\int _0^{t}\omega _k(t')dt'\mp \omega _0(t)\int _0^{r}\frac{dr}{\sqrt{AB}},~~~~k=1,2,3....
\end{equation}
Thus from equation (16), using the solutions (26) and (27) into equation (18), the wave functions for absorption and emmission of scalar particle can be expressed as
\begin{equation}
\psi _{emm.}(r,t)=exp\left\lbrace -\frac{i}{\hbar}\left[\left(\int _0^t \omega _0(t')dt'+\Sigma _k \hbar ^k \int _0^t \omega _k (t')dt'\right)-(\omega _0(t)+\Sigma _k \hbar ^k \omega _k (t))\int _0^r \frac{dr}{\sqrt{AB}}\right]\right\rbrace
\end{equation}
and
\begin{equation}
\psi _{abs.}(r,t)=exp\left\lbrace -\frac{i}{\hbar}\left[\left(\int _0^t \omega _0(t')dt'+\Sigma _k \hbar ^k \int _0^t \omega _k (t')dt'\right)+(\omega _0(t)+\Sigma _k \hbar ^k \omega _k (t))\int _0^r \frac{dr}{\sqrt{AB}}\right]\right\rbrace
\end{equation}
respectively. Due to tunnelling across the horizon, there will be a change of sign of the metric coefficients in the $(r,t)$-part of the metric and as a result, function of $t$ coordinate has an imaginary part which will contribute to the probabilities. So we write
\begin{equation}
P_{abs.}=|\psi _{abs.}(r,t)|^2=exp\left\lbrace \frac{2Im}{\hbar}\left[\left(\int _0^t \omega _0(t')dt'+\Sigma _k \hbar ^k \int _0^t \omega _k (t')dt'\right)+(\omega _0(t)+\Sigma _k \hbar ^k \omega _k (t))\int _0^r \frac{dr}{\sqrt{AB}}\right]\right\rbrace
\end{equation}
and 
\begin{equation}
P_{emm.}=|\psi _{emm.}(r,t)|^2=exp\left\lbrace \frac{2Im}{\hbar}\left[\left(\int _0^t \omega _0(t')dt'+\Sigma _k \hbar ^k \int _0^t \omega _k (t')dt'\right)-(\omega _0(t)+\Sigma _k \hbar ^k \omega _k (t))\int _0^r \frac{dr}{\sqrt{AB}}\right]\right\rbrace .
\end{equation}
To have some simplification, we shall now use the physical fact that all incoming particles certainly cross the horizon, {\it i.e.,} $P_{abs.}=1$. So from equation (30),
\begin{equation}
Im~\left(\int _0^t \omega _0(t')dt'+\Sigma _k \hbar ^k \int _0^t \omega _k (t')dt'\right)=-Im~(\omega _0(t)+\Sigma _k \hbar ^k \omega _k (t))\int _0^r \frac{dr}{\sqrt{AB}}
\end{equation}
and hence $P_{emm.}$ simplifies to 
\begin{equation}
P_{emm.}=exp\left\lbrace -\frac{4}{\hbar}(\omega _0(t)+\Sigma _k \hbar ^k \omega _k (t))Im\int _0^r \frac{dr}{\sqrt{AB}}\right\rbrace .
\end{equation}
Then form the principle of "detailed balance" [10,11,12] (which states states transitions between any two states take place with equal frequency in either direction at equilibrium), we write
\begin{equation}
P_{emm.}=exp\left\lbrace -\frac{\omega _0(t)}{T_h}\right\rbrace P_{in}=exp\left\lbrace -\frac{\omega _0(t)}{T_h}\right\rbrace .
\end{equation}
So, comparing (33) and (34), the of the BH is given by
\begin{equation}
T_h=\frac{\hbar}{4}\left[1+\Sigma _k \hbar ^k \frac{\omega _k(t)}{\omega _0(t)}\right]^{-1}\left[Im\int _0^r \frac{dr}{\sqrt{AB}}\right]^{-1},
\end{equation}
where
\begin{equation}
T_h=\frac{\hbar}{4}\left[Im\int _0^r \frac{dr}{\sqrt{AB}}\right]^{-1}
\end{equation}
is the usual Hawking temperature of the BH. Thus, due to quantum corrections, the temperature of the BH is modified from the Hawking temperature and both the temperatures are functions of $t$ and $r$. Note that equation (36) is the standard expression for semi-classical Hawking temperature and it is valid for non-spherical metric also. However, for spherical metric, one can use the Taylor series expansions (7) and (8) near the horizon and obtain $T_H$ as given in equation (12) by performing the contour integration. The ambiguity of factor of two (as mentioned earlier) in the Hawking temperature does not arise here.

Further, one may note that solutions (26) or (27) are the unique solutions to the equations (19) or (22) except for a pre-multiplication factor. This arbitrary multiplicative factor does not appear in the expression for Hawking temperature, only the particle energy ($\omega _0$) or $\omega _k$ are re-scaled. As quantum correction term contains $\frac{\omega _k}{\omega _0}$, so it does not involve the arbitrary multiplicative factor and hence unique.

To have some interpretation about the arbitrary functions $\omega _k(t)$ appearing in the quantum correction terms, we make use of dimensional analysis. As $S_0$ has the dimension $\hbar$, so the arbitrary function $\omega _k(t)$ has the dimension $\hbar ^{-k}$. In standard choice of units, namely $G=c=K_B=1$, $\hbar \sim M_p^2$ and so $\omega _k \sim M^{-2k}$, where $M$ is the mass of the BH.

Similar to the Hawking temperature, the surface gravity of the BH is modified due to quantum corrections. If $\kappa _c$ is the semi-classical surface gravity corresponding to Hawking temperature, {\it i.e.,} $\kappa _c=2\pi T_H$, then the quantum corrected surface gravity $\kappa =2\pi T_H$ is related to the semi-classical value by the relation
\begin{equation}
\kappa =\kappa _c\left[1+\Sigma _k \hbar ^k \frac{\omega _k(t)}{\omega _0(t)}\right]^{-1}.
\end{equation} 
Moreover, based on the dimensional analysis, if we choose for simplicity,
\begin{equation}
\omega _k(t)=\frac{a^k \omega _0(t)}{M^{2k}}~,~~~'a'~is~a~dimensionless~parameter,
\end{equation}
then the expression (37) is simplified to 
\begin{equation}
\kappa =\kappa _0 \left(1-\frac{\hbar a}{M^2}\right)^{-1}.
\end{equation}
This is related to the one loop back reaction effects in the spacetime [6,30] with the parameter $a$ corresponding to trace anomaly. Higher order loop corrections to the surface gravity can be obtained similarly by suitable choice of the functions $\omega _k(t)$. For static BHs, Banerjee {\it et al.} [19] have studied these corrections in details. Lastly, it is worthy to mention that identical result for BH temperature may be obtained if we use the Painleve coordinate system as in the previous section.

\section{Entropy Function and Quantum Correction}

We shall now examine how the semi-classical Bekenstein-Hawking area law, namely $S_{BH}=\frac{A}{4\hbar}$ ($A$ is the area of the horizon) is modified due to quantum corrections described in the previous section. The first law of the BH mechanics which is essentially the energy conservation relation, related the change of BH mass ($M$) to the change of its entropy ($S_{BH}$), electric charge ($Q$) and angular momentum ($J$) as
\begin{equation}
dM=T_h dS_{BH}+\Phi dQ+\Omega dJ.
\end{equation}
Here, $\Omega$ is the angular velocity and $\Phi$ is the electrostatic potential. So, for non-rotating uncharged BHs, the entropy has the simple form
\begin{equation}
S_{BH}=\int \frac{dM}{T_h},
\end{equation}
or using equation (35) for $T_h$, we get
\begin{equation}
S_{BH}=\int \left[1+\Sigma _k \hbar ^k \frac{\omega _k(t)}{\omega _0(t)}\right]\frac{dM}{T_H}.
\end{equation}
For the choice (38) corresponding to one loop back reaction effects, we have from (42), the quantum corrected BH entropy as
\begin{equation}
S_{BH}=\int \left[1+\frac{a\hbar}{M}+\frac{a^2 \hbar ^2}{M^2}+ .....\right]\frac{dM}{T_H}.
\end{equation}
The first term is the usual semi-classical Bekenstein-Hawking entropy and the subsequent terms are the quantum corrections of different order. For static BHs, Banerjee {\it et al.} [19] have shown the correction terms of which the leading one gives the standard logarithmic correction. On the other hand, for non-static BHs, as the proportionality factors are time-dependent and arbitrary ({\it see equation} (42)) so the leading order correction term may not be logarithmic. For future work, we shall attempt to determine physical interpretation of the arbitrary time-dependent proportionality factors so that quantum corrections may be evaluated.

\section{Hamilton-Jacobi Method for Massive Particles: Quantum Corrections}

The KG equation for a scalar field $\psi$ describing a scalar particle of mass $m_0$ has the form [10]
\begin{equation}
\left(\fbox{}+\frac{{m_0}^2}{\hbar ^2}\right)\psi =0,
\end{equation}
where the box operator '$\fbox{}$' is evaluated in the background of a general static BH metric of the form
\begin{equation}
ds^2=-A(r)dt^2+\frac{dr^2}{B(r)}+r^2d{\Omega _2}^2.
\end{equation}
The explicit form of the KG equation for the metric (45) is 
\begin{equation}
-\frac{1}{A}\frac{\partial ^2 \psi}{\partial t^2}+B\frac{\partial ^2 \psi}{\partial r^2}+\frac{1}{2A}\frac{\partial (AB)}{\partial r}\frac{\partial \psi}{\partial r}+\frac{2B}{r}\frac{\partial \psi}{\partial r}+\frac{1}{r^2 sin~\theta}\frac{\partial}{\partial \theta}\left(sin~\theta \frac{\partial \psi}{\partial \theta}\right)+\frac{1}{r^2 sin^2\theta}\frac{\partial ^2 \psi}{\partial \phi ^2}=\frac{{m_0}^2}{\hbar ^2}\psi (t,r,\theta ,\phi).
\end{equation}
Due to spherical symmetry, we can decompose $\phi$ in the form
\begin{equation}
\psi (t,r,\theta ,\phi)=\Phi (t,r)Y_l^{m}(\theta ,\phi),
\end{equation}
where $\phi$ satisfies [10]
\begin{equation}
\frac{1}{A}\frac{\partial ^2 \psi}{\partial t^2}-B\frac{\partial ^2 \psi}{\partial r^2}-\frac{1}{2A}\frac{\partial (AB)}{\partial r}\frac{\partial \psi}{\partial r}-\frac{2B}{r}\frac{\partial \psi}{\partial r}+\left\lbrace \frac{l(l+1)}{r^2}+\frac{{m_0}^2}{\hbar ^2}\right\rbrace \Phi (t,r)=0.
\end{equation}
If we substitute the standard ansatz for the semi-classical wave function, namely
\begin{equation}
\phi (t,r)=exp\left\lbrace -\frac{i}{\hbar}S(r,t)\right\rbrace ,
\end{equation}
then the action $S$ will satisfy the following differential equation:
\begin{equation}
\left[\frac{1}{A}{\left(\frac{\partial S}{\partial t}\right)}^2-B{\left(\frac{\partial S}{\partial r}\right)}^2-E_0^{2}(r)\right]-\frac{\hbar}{i}\left[\frac{1}{A}\frac{\partial ^2 S}{\partial t^2}-B^2\frac{\partial ^2 S}{\partial r^2}-\left\lbrace \frac{1}{2A}\frac{\partial (AB)}{\partial r}+\frac{2B}{r}\right\rbrace \frac{\partial S}{\partial r}\right]=0,
\end{equation}
where $E_0^{2}=m_0^{2}+\frac{L^2}{r^2}$ and $L^2=l(l+1)\hbar ^2$ is the angular momentum. To incorporate quantum corrections over the semi-classical action, we expand the actions in powers of Planck constant $\hbar$ as
\begin{equation}
S(r,t)=S_0(r,t)+\Sigma _k \hbar ^kS_k(r,t),
\end{equation}
where $S_0$ is the semi-classical action and $k$ is a positive integer. Now substituting this ansatz for $S$ in the differential equation (50) and equating different powers of $\hbar$ on both sides, we obtain the following set of partial differential equations:
\begin{equation}
\hbar ^0: \frac{1}{A}{\left(\frac{\partial S}{\partial t}\right)}^2-B{\left(\frac{\partial S}{\partial r}\right)}^2-E_0^{2}(r)=0,
\end{equation}
\begin{equation}
\hbar ^1: \frac{2}{A}\frac{\partial S_0}{\partial t}\frac{\partial S_1}{\partial t}-2B\frac{\partial S_0}{\partial r}\frac{\partial S_1}{\partial r}-\frac{1}{i}\left[\frac{1}{A}\frac{\partial ^2 S_0}{\partial t^2}-B^2\frac{\partial ^2 S_0}{\partial r^2}-\left\lbrace \frac{1}{2A}\frac{\partial (AB)}{\partial r}+\frac{2B}{r}\right\rbrace \frac{\partial S_0}{\partial r}\right]=0,
\end{equation}
\begin{equation}
\hbar ^2: \frac{1}{A}{\left(\frac{\partial S_1}{\partial t}\right)}^2+\frac{2}{A}\frac{\partial S_0}{\partial t}\frac{\partial S_2}{\partial t}-B{\left(\frac{\partial S_1}{\partial r}\right)}^2-2B\frac{\partial S_0}{\partial r}\frac{\partial S_2}{\partial r}-\frac{1}{i}\left[\frac{1}{A}\frac{\partial ^2 S_1}{\partial t^2}-B^2\frac{\partial ^2 S_1}{\partial r^2}-\left\lbrace \frac{1}{2A}\frac{\partial (AB)}{\partial r}+\frac{2B}{r}\right\rbrace \frac{\partial S_1}{\partial r}\right]=0,
\end{equation}
and so on.

To solve the semi-classical action $S_0$, we start with the standard separable choice [10]
\begin{equation}
S_0(r,t)=\omega _0 t+D_0(r).
\end{equation}
Substituting this choice in equation (52), we obtain
\begin{equation}
D_0=\pm \int _0^{r}\sqrt{\frac{\omega _0^{2}-AE_0^{2}}{AB}}dr=\pm I_0~~~(say),
\end{equation}
where + or - sign corresponds to absorption or emmission of scalar particle. Now substituting this choice for $S_0$ in equation (53), we have the differential equation for first order corrections $S_1$ as
\begin{equation}
\frac{\partial S_1}{\partial t}\mp \sqrt{AB}\sqrt{1-\frac{AE_0^{2}}{\omega _0^{2}}}\frac{\partial S_1}{\partial r}\mp \frac{\sqrt{AB}}{i}\left[-\frac{1}{r}\sqrt{1-\frac{AE_0^{2}}{\omega ^{2}}}+\frac{\frac{\partial A}{\partial r}\frac{E_0^{2}}{\omega ^2}-\frac{2AL^2}{\omega _0^{2}r^3}}{4\sqrt{1-\frac{AE_0^{2}}{\omega ^2}}}\right]=0.
\end{equation}
As before, $S_1$ can be written in separable form as
\begin{equation}
S_1=\omega _1 t+D_1(r),
\end{equation}
where
\begin{equation}
D_1=\int _0^{r} \frac{dr}{\sqrt{AB}\sqrt{1-\frac{AE_0^{2}}{\omega _0^{2}}}}\left[\pm \omega _1-\frac{\sqrt{AB}}{i}\left\lbrace -\frac{1}{r}\sqrt{1-\frac{AE_0^{2}}{\omega ^{2}}}+\frac{\frac{\partial A}{\partial r}\frac{E_0^{2}}{\omega ^2}-\frac{2AL^2}{\omega _0^{2}r^3}}{4\sqrt{1-\frac{AE_0^{2}}{\omega ^2}}}\right\rbrace \right]=\pm I_1-I_2.
\end{equation}
Now due to complicated form, if we retain terms upto first order quantum corrections, {\it i.e.,}
\begin{equation}
S=S_0+\hbar S_1=(\omega _0+\hbar \omega _1)t+\lbrace{D_0+\hbar D_1(r)\rbrace},
\end{equation}
then the wave function denoting absorption and emmission solutions of the KG equation (48) using (49), are of the form
\begin{equation}
\phi _{abs.}=exp\lbrace{-\frac{i}{\hbar}(\overline{\omega _0+\hbar \omega _1} t+\overline{I_0+\hbar I_1-\hbar I_2})\rbrace}
\end{equation}
and 
\begin{equation}
\phi _{emm.}=exp\lbrace{-\frac{i}{\hbar}(\overline{\omega _0+\hbar \omega _1} t-\overline{I_0+\hbar I_1-\hbar I_2})\rbrace}.
\end{equation}
It is to be noted that in course of tunnelling across the horizon, the coordinate nature changes, {\it i.e.,} more precisely the sign of the metric coefficients in the ($r$,$t$)-hyperplane are altered. Thus, we can interpret as that the time coordinate has an imaginary part in crossing the horizon and accordingly the temporal part has contribution to the probabilities [19,31]. Thus absorption and emmission probabilities are given by
\begin{equation}
P_{abs.}=|\phi _{in}|^2=exp\left\lbrace \frac{2}{\hbar}(Im~\overline{\omega _0+\hbar \omega _1} t)+Im~\overline{I_0+\hbar I_1}-Im~\hbar I_2\right\rbrace
\end{equation}
and 
\begin{equation}
P_{emm.}=|\phi _{out}|^2=exp\left\lbrace -\frac{i}{\hbar}(Im~\overline{\omega _0+\hbar \omega _1} t)-Im~\overline{I_0+\hbar I_1}-Im~\hbar I_2\right\rbrace .
\end{equation}
In the classical limit $\hbar \rightarrow 0$, there is no reflection, so all ingoing particles should be absorbed and hence [31]
\begin{equation}
lim_{\hbar \rightarrow 0}P_{abs.}=1.
\end{equation} 
So, from equation (63), we must have
\begin{equation}
Im~\omega _0 t=Im~I_0~~~~~~Im~(\omega _1 t-I_2)=Im~I_1
\end{equation}
and as a result $P_{emm.}$ simplifies to
\begin{equation}
P_{emm.}=exp \left[-\frac{4\omega _0}{\hbar}Im~\left\lbrace \int _0^{r}\frac{dr}{\sqrt{AB}}\left(\sqrt{1-\frac{AE_0^2}{\omega _0^2}}+\frac{\hbar (\frac{\omega _1}{\omega _0})}{\sqrt{1-\frac{AE_0^2}{\omega _0^2}}}\right)\right\rbrace \right].
\end{equation}
Using the principle of "detailed balance" [10,11,20,21], namely
\begin{equation}
P_{emm.}=exp\left\lbrace -\frac{E}{T_h}\right\rbrace P_{in}=exp\left\lbrace -\frac{E}{T_h}\right\rbrace ,
\end{equation}
the temperature of the BH is given by
\begin{equation}
T_h=\frac{\hbar E}{4\omega _0}\left[Im\left\lbrace \int _0^{r}\frac{dr}{\sqrt{AB}}\left(\sqrt{1-\frac{AE_0^2}{\omega _0^2}}+\frac{\hbar (\frac{\omega _1}{\omega _0})}{\sqrt{1-\frac{AE_0^2}{\omega _0^2}}}\right)\right\rbrace \right]^{-1},
\end{equation}
where the semi-classical Hawking temperature of the BH has the expression
\begin{equation}
T_H=\frac{\hbar E}{4\omega _0}\left[Im\int _0^{r}\frac{dr}{\sqrt{AB}}\sqrt{1-\frac{AE_0^2}{\omega _0^2}}\right]^{-1}.
\end{equation}
Now, to obtain the modified form of the surface gravity of the BH, we start with the usual relation between surface gravity and Hawking temperature, namely
\begin{equation}
\kappa _H=2\pi T_H,
\end{equation}
where $T_H$ is given by equation (70).\\
So the quantum corrected surface gravity is given by
\begin{equation}
\kappa _{QC}=2\pi T_h.
\end{equation}
Further, for the present non-rotating, uncharged, static BHs, using the law of BH thermodynamics $dM=T_h dS$, we have the expression for the entropy of the BH as
\begin{equation}
S_{BH}=\int \frac{4\omega _0}{\hbar E}\left(1+\frac{\hbar \omega _1}{\omega _0}\right)dM \int _0^{r} \frac{dr}{\sqrt{AB}}.
\end{equation}
Finally, it is easy to see from equation (69) that near the horizon the presence of $E_0^{2}$ term can be neglected as it is multiplied by the metric coefficient $A$. Therefore, the quantum corrected (upto first order) temperature of the BH (in equation (69)) reduces to
\begin{equation}
T_h=\frac{\hbar E}{4\omega _0}\left(1+\frac{\hbar \omega _1}{\omega _0}\right)^{-1}\left[\int _0^{r}\frac{dr}{\sqrt{AB}}\right]^{-1}
\end{equation}
and the Hawking temperature (given in equation (70)) becomes
\begin{equation}
T_H=\frac{\hbar E}{4\omega _0}\left[\int _0^{r}\frac{dr}{\sqrt{AB}}\right]^{-1}.
\end{equation}
So we have
\begin{equation}
T_h=\left(1+\frac{\hbar \omega _1}{\omega _0}\right)^{-1} T_H.
\end{equation}
We see that if the energy of the tunnelling particle is chosen as $\omega _0$ ({\it i.e.,} $E=\omega _0$) and $\omega _1=\frac{\beta _1}{M}$ (for notations see {\it Banerjee et al.} [19]) then the Hawking temperature given by equation (75) is the usual one derived for massless particles and the quantum corrected temperature $T_h$ given in equation (76) agrees with that of Banerjee {\it et al.} [19] for massless particle. Therefore, Hawking temperature near the horizon remains same for both massless and non-zero mass tunnelling particle and it agrees with the claim of Srinivasan {\it et al.} [10] and Banerjee {\it et al.} [19]. For future work, it will be interesting to calculate the temperature of the BH for tunnelling non-zero mass particle with full quantum corrections and examine whether the result agrees with that of Banerjee {\it et al.} [19] near the horizon. Finally, it will be nice to determine quantum corrected entropy of the BH in a convinient form.

\section{SUMMARY OF THE WORK}

This work is an attempt to study quantum corrections to Hawking radiation of massless particle from a dynamical BH as well as for massive particle from a static BH. At first, radial null geodesic tunnelling approach has been used with Painleve-type choice of coordinate system to derive semi-classical Hawking temperature. Then a full quantum mechanical calculations have been performed writing action in a power series of the Planck constant $\hbar$ to evaluate the quantum corrections to the Hawking temperature. Subsequently, quantum corrected surface gravity has been calculated and it is found that one loop back reaction effects in the spacetime can be obtained by suitable choice of the arbitrary functions and parameters. Finally, an expression for the quantum corrected entropy of the BH has been evaluated. It is found that due to presence of the arbitrary functions in the expression for entropy, the leading order quantum correction may not be logarithmic in nature. On the other hand, in case of Hawking radiation of massive particle from static BH, it is found that Hawking temperature near the horizon does not depend on the mass term as predicted by Srinivasan {\it et al.} [10] and Banerjee {\it et al.} [16-18]. For future work, we shall try to find a solution of the partial differential equation (19) in a more simpler form so that more physical interpretations can be done from the BH parameters. Also, it will be interesting to calculate temperature of the BH for tunnelling non-zero mass particle with full quantum correction and examine whether the result agrees with that of Banerjee {\it et al.} [19] near the horizon. Finally, it will be nice to determine quantum corrected entropy of the BH in a convinient form. 

\begin{acknowledgments}

The authors are thankful to IUCAA, Pune, India for their warm hospitality and research facilities as the work has been done there during a visit. Also SC acknowledges the UGC-DRS Programme in the Department of Mathematics, Jadavpur University. The author SS is thankful to UGC-BSR Programme of Jadavpur University for awarding research fellowship.

\end{acknowledgments}

\end{document}